\documentclass[prl,twocolumn,showpacs,superscriptaddress]{revtex4}
\usepackage{graphicx}
\usepackage{dcolumn}
\usepackage{bm}
\begin{document}
\title{Checkerboard and stripe inhomogeneities in cuprates}

\author{G. Seibold}
\affiliation{Institut f\"ur Physik, BTU Cottbus, PBox 101344,
         03013 Cottbus, Germany}
\author{J. Lorenzana}
\affiliation{SMC-INFM,ISC-CNR, Dipartimento di Fisica,
Universit\`a di Roma La Sapienza, P. Aldo Moro 2, 00185 Roma, Italy}
\author{M. Grilli}
\affiliation{INFM-SMC Center and Dipartimento di Fisica, 
Universit\`a di Roma "La Sapienza" piazzale Aldo Moro 5, I-00185 Roma, Italy}
\begin{abstract}
We systematically investigate charge-ordering phases by means of a restricted
and unrestricted Gutzwiller approximation to the single-band Hubbard
model with nearest ($t$) and next-nearest neighbor hopping ($t'$). 
When  $|t'/t|$ is small, as appropriate
for ${\rm La_{2-x}Sr_xCuO_4}$, stripes are found, whereas in compounds 
with larger $|t'/t|$ (such as ${\rm Ca_{2-x}Na_x CuO_2Cl_2}$ and ${\rm
 Bi_2Sr_2CaCu_2O_{8+\delta}}$)  checkerboard structures are favored.
In contrast to the linear doping dependence found for stripes
the charge periodicity of checkerboard textures is locked to
$4$ unit cells over a wide doping range. 
In addition we find that checkerboard structures are favored at
surfaces. 
\end{abstract}
\date{\today}
\maketitle

%%%%%%%%%%%%%%%%%%%%%%%%%%%%%%%%%%%%%%%%%%%%%%%%%%%%%%%%%%%%%%%%%%%%%%
The presence of charge and related spin inhomogeneities in underdoped
high-temperature superconducting cuprates has received substantial
experimental support\cite{tra95,yam98,moo00nat,abb05}. Less clear is the
symmetry of the underlying textures, an issue which is 
strongly debated\cite{tra98,moo00nat,hin04,fin04}.

Neutron scattering  experiments on lanthanum cuprates (LCO) 
are usually interpreted in terms of one-dimensional 
modulations (stripes) where the two-dimensional symmetry of the 
scattering is explained with the presence of orthogonal
stripe domains\cite{tra98}.
Indeed, a substantial anisotropy in the {\em dynamic} scattering between the two planar
axis, as expected for fluctuating stripes, has been explicitely 
demonstrated in studies on untwinned samples of 
YBa$_2$Cu$_3$O$_y$ (YBCO)\cite{moo00nat,hin04}.
On the other hand, scanning tunneling microscopy (STM) in other classes  of
cuprates, namely 
the bilayer ${\rm Bi_2Sr_2CaCu_2O_{8+\delta}}$ (Bi-2212)\cite{hof02sci,how03,ver04,mce05} 
and the single-layer ${\rm Ca_{2-x}Na_x CuO_2Cl_2}$ (Na-CCOC)\cite{han04} 
has revealed two-dimensional modulations (checkerboards).
The peculiar characteristics of the Fermi surface 
of Na-CCOC \cite{she05} is also in agreement with the expected features of a
disordered checkerboard phase\cite{sei00}.

Charge-ordering (CO) phenomena thus seem to be  common (and possibly
generically present) in cuprates but the different symmetries found are
rather puzzling and makes one wonder if one should not 
reinterpret the neutron scattering experiments in LCO and YBCO,
in terms of two-dimensional textures also\cite{fin04}. 

CO where predicted in cuprates before any experimental 
detection\cite{cotheory}.
However, its cooperation or competition with superconductivity is still
debated. 
In this regard 
a possible direct  relation between $T_c$ and the intensity of incommensurate
low-energy scattering \cite{wak04wak99} 
demands for a deeper understanding of the 
physical mechanisms inducing CO and of the symmetry of the textures.

In this work we address this issue on the basis of the 
one-band Hubbard Hamiltonian.
It has been argued that within this model, the ratio between
nearest- ($t$) and next-nearest-
 ($ t'$) neighbor hopping $t'/t<0$ is the 
main electronic parameter 
characterizing the different cuprate families\cite{pav01,tan04}. 
We find a transition from one-dimensional to two-dimensional 
textures as $|t'/t|$ is increased. In addition, we find that the
tendency to form checkerboard structures is increased at surfaces. 
Our results provide a clue to interpret the
contrasting experimental results regarding the symmetry of CO structures
in different cuprate compounds. 

We solve the Hubbard model within a unrestricted 
Gutzwiller approximation (GA). The GA energy is minimized with
respect to unconstrained charge, spin and double occupancy
distributions (for details see Ref.~\cite{sei98}).
Generally we restrict to solutions without 
spin canting but we have checked the results  lifting this restriction
in several cases. In the case of stripes, once the unrestricted
solution has been found, the symmetry of the solution has been imposed  
on a supercell and the problem has been solved in momentum space,
allowing to obtain smooth curves of the energy as a function of
doping. In the other cases, the curves are not smooth due to 
restricted sampling but, we estimate, that the finite size 
corrections to the energy of single points are negligible due to the
large size of the cells (up to $16\times16$).

The use of the present mean-field-like scheme  
to address the phenomenon of CO in
cuprates requires further justification:
{\it i}) Previous studies have shown that the GA applied to a three-band
Hubbard model and in the present model captures the phenomenology of 
stripes in LCO cuprates\cite{lor02bsei04a}.
 In particular
it accounts for 
the behavior of  incommensurability and of the chemical potential
as a function of doping. This requires an accurate evaluation of the 
small energy difference between different stripe solutions.   
{\it ii)} Response functions, due to fluctuations on top of 
these mean-field states, are in good agreement 
with spectroscopies\cite{tra04,uch91}, both in the charge\cite{lor03} 
and in the spin sector\cite{sei05lor05sei06}. 
{\it iii}) The accuracy of the 
charge distributions can be directly tested comparing with exact methods. 
We have found that the charge profile in the GA is indistinguishable
from density matrix renormalization group computations\cite{whi03} 
revealing stripes on a $7\times 6$ Hubbard cluster doped
with 4 holes. 
{\it iv}) Finally, the charge profile predicted within the GA 
in the three-band Hubbard model\cite{lor02bsei04a}, has recently been found to be in
excellent agreement with a charge sensitive measurement\cite{abb05}.

All this establishes the GA as a very reliable technique
to determine charge inhomogeneities in a strongly correlated system.
Although, as any mean-field-like method, it cannot be trusted to determine
the presence of long-range order (especially in the delicate spin-sector)
it is expected to give a reasonably accurate estimate of the
relative ground state energies of different textures, which is what we
need here, and of the short-range correlations which
determine it.

For LCO we fix the Hubbard on-site repulsion $U/t=8$ and
$t'/t=-0.2$. This parameter set
and a very similar one reproduce  the 
spectrum of magnetic excitations both in the undoped and doped 
phases\cite{sei05lor05sei06}.
The Hubbard interaction $U$ and the nearest-neighbor hopping 
are not expected to depend significantly on the material
given the similarity of superexchange interactions\cite{tan04}. 
Instead $t'$ has been found to be quite sensitive to the 
cuprate family\cite{tan04,pav01}. 
Photoemission studies in the insulating parent compounds have shown
that there is substantial splitting among states with momentum
$(\pi/2,\pi/2)$ and $(\pi,0)$ (we set the lattice constant $a\equiv
1$). Detailed studies of a single hole in one band
models suggest that this splitting roughly scales with
$t'$\cite{tan04}.  
Taking $t'/t=-0.2$ for LCO\cite{sei05lor05sei06}
we estimate $t'$ in Bi-2212 and CCOC by
rescaling $t'$ with the observed splitting\cite{tan04}. In this way 
we obtain $t'\sim -0.4t$ for  Bi-2212  and $t' \sim -0.5 t$  for CCOC.  
The increase of $|t'/t|$ from LCO to Bi-2212 is in agreement with
LDA results \cite{pav01} whereas for CCOC
the present semiempirical estimate is significantly larger.

%%%%%%%%%%%%%%%%%%%%%%%%%%%%%%%%%%%%%%%%%%%%%%%%%%%%%%%%%%%%%%%%%%%%%%
\begin{figure}
%\hspace{4 truecm}
\includegraphics[width=8cm,clip=true]{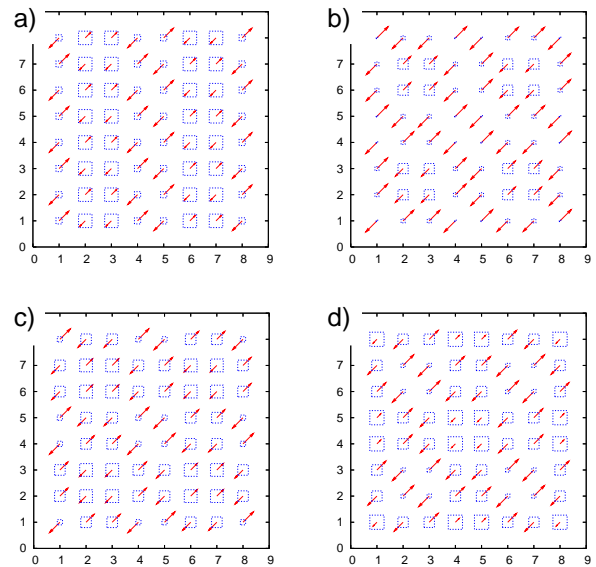}
\caption{(Color online) CO textures for $U/t=8$. The length of arrows is
  proportional to the spin density whereas the squares are
  proportional to the hole density minus the hole density in the
  insulator.  
(a) Bond-Centered stripe texture with charge spacing $d=4$,  doping $n_h=1/8$,
 and $t'/t=-0.2$; (b) SF checkerboard with $d=4$ at $x=0.06$, and $t'/t=-0.5$;
(c) same as (b) with $n_h=1/8$; (d) LF checkerboard with $d=4$ and $t'/t=-0.5$
at $n_h=1/8$
}
\label{textures}
%
%       fig. 1: 
\end{figure}
%%%%%%%%%%%%%%%%%%%%%%%%%%%%%%%%%%%%%%%%%%%%%%%%%%%%%%%%%%%%%%%%%%%%%%%%
Fig.~\ref{textures} shows the most relevant textures
found in this study. (a) Is the bond-centered (BC) stripe 
solution already reported in
previous studies\cite{lor02bsei04a}. (b)-(d) are checkerboard textures. 
At low doping ($n_h$) and for large values of $|t'/t|$ we find the configuration  
(b) to be particularly stable. It can be seen as a lattice of
ferromagnetic (FM) polarons or ``ferrons''  and is refered to as small ferron
checkerboard (SFC). Each ferron consists of a $2\times2$ plaquette with uniform
magnetization and resides at the intersections of a crossed array
of antiphase domain walls of the antiferromagnetic (AFM) order parameter. 
As doping increases the domain walls populate more uniformly and  
evolve to a configuration that resembles a crossed array of stripes (c). 
At higher doping the size of the ferromagnetic islands increases, producing a  
large ferron checkerboard (LFC) structure (d). In addition, we have considered
arrays of spin-polarons (not shown) consisting of single spin flips 
which locally bind the holes.

Fourier transforming the charge and spin distributions, we find that the most
intense Bragg peaks for the  $d=4$ checkerboards (SFC,LFC) 
are at ${\bf Q}_c=(\pm \pi/2,0),(0,\pm \pi/2)$ (charge) and 
at ${\bf Q}_s=(\pi\pm\pi/4,\pi\pm\pi/4)$ (spin). As
argued by Tranquada\cite{tra98} the rotation between ${\bf Q}_c$ and
${\bf Q}_s$ is a signature for a checkerboard
phase since in case of stripes one would have   ${\bf Q}_c=(0\pm 2\pi/d,0)$
and ${\bf Q}_s=  (\pi\pm\pi/d,0)$
or ${\bf Q}_c=(0,0\pm 2\pi/d)$ and ${\bf Q}_s=  (0,\pi\pm\pi/d)$ depending on orientation.

\begin{figure}
%\vspace*{-2cm}
\includegraphics[width=8cm,clip=true]{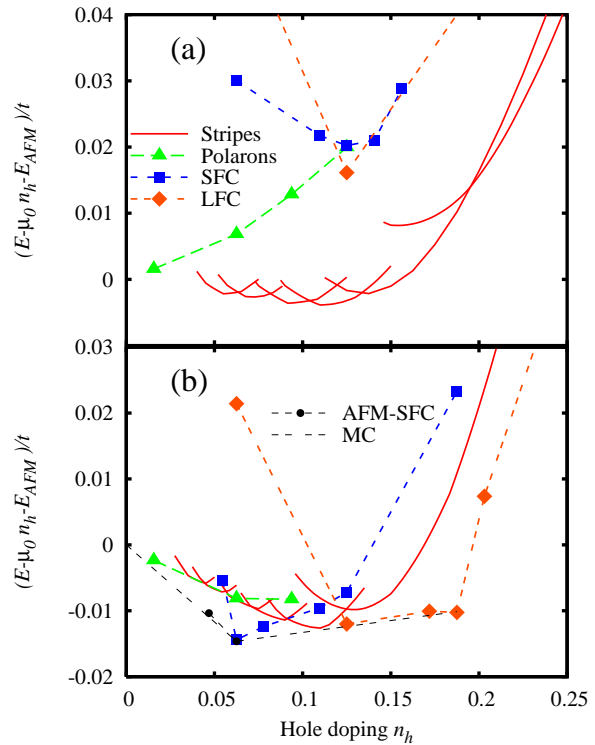}
\caption{(Color online) Energy per site as a function of doping for
  the different textures studied and for $t'/t=-0.2$ (a) and $t'/t=-0.5$ (b). 
For clarity we subtracted the energy of the AFM
  solution and the line $\mu_0 n_h$ with $\mu_0=-1.6t$.
Different  choices of $\mu_0$ correspond to different choices of the origin of
the  energy of the single particle states and do not change the physics.
 In each curve for the stripes the charge
  periodicity perpendicular to the stripe is fixed and takes the 
values (from left to right) $d=10,8,6,5,4,3$. In panel (b) we also
  show the energies allowing for spin
  canting and AFM-SFC mixing (filled circles) 
and  Maxwell construction (MC) for AFM-SFC phase
  separation and for SFC and LFC phase separation.
 }
\label{edx}
%
%       fig. 2: 
\end{figure}

In Fig.~\ref{edx} 
we report the values of the energy per site $E$
for selected checkerboard (dashed lines) and stripe (solid lines) 
textures in case
of $t'/t = -0.2$ (a) and $t'/t = -0.5$ (b). 
 The main general conclusion one can draw from Fig.~\ref{edx}
is that a sizable value of the next-nearest-neighbor hopping $t'$ 
stabilizes checkerboard solutions with respect to stripes, which
are more stable at small $|t'/t|$.
This is the most
generic and relevant finding of this work and provides a rational on why
stripes are observed in LCO materials (where $t'/t \approx -0.2$), while 
STM experiments detect checkerboard structures in Na-CCOC and
Bi-2212, where $t'/t$ is known to be substantially larger\cite{tan04}.

\begin{figure}
\includegraphics[width=8cm,clip=true]{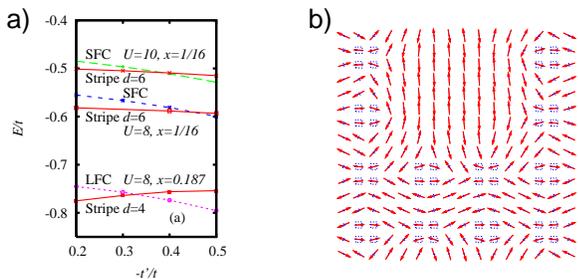}
\caption{(Color online) (a) Energy per site vs. $-t'/t$ for $U/t=8,10$ for stripes and
  checkerboard at $n_h=0.0625,0.187$. (b)Mixed AFM-SFC solution at 
  $n_h= 0.047$ and allowing for spin canting. }
\label{spincanting}
%
%       fig. 5: 
\end{figure}

Fig.~\ref{spincanting}(a)
details on this issue by comparing the energy of checkerboard
with stripe solutions as a function of $t'/t$. Although most of the CO 
structures gain energy by increasing $|t'|$, 
it is clearly apparent that the checkerboard
solutions  take a much greater advantage
from this increase and become more stable at $|t'/t|>0.35\sim 0.45$
depending on doping.

Bond-centered checkerboards take advantage from the ``ferronic'' nodes of the charge 
texture, while site-centered checkerboards (not shown) lack this important feature
and have much higher energies. This is a major difference with respect
to stripe structures, where bond-centered and site-centered textures
are nearly degenerate. Another important difference lies in the doping dependence
of the periodicity.
The most stable stripe solution at low doping (cf. Fig.~\ref{edx})  
has $d\sim 1/(2 n_h)$ (with $d\approx 4-10$)
providing an explanation\cite{lor02bsei04a} for the 
well-known linear behavior between
doping and incommensurability\cite{yam98}.  In contrast, 
the bond-centered checkerboards lose (magnetic)
energy in the domain walls and it is more convenient for them
to adjust the charge in the FM plaquettes keeping the domain walls as short as
possible. Indeed the checkerboard structures with $d > 4$ (not shown)
are at higher energy in the whole doping range.
 This provides an explanation for the fact that in Na-CCOC the
charge periodicity is independent of doping and locked at $d=4$\cite{han04}. 
For Bi-2212 
$d \approx 4-4.7$ \cite{how03,ver04,mce05} has been reported in STM experiments. 
However, here the interpretation is more difficult due to the presence of lattice
modulations, mesoscopic inhomogeneities, and less neat CO peaks.

The SFC is particularly stable for large $|t'/t|$ at $n_h=1/16\sim
0.06$ (cf. Fig.~\ref{edx}b) when each ferron accommodates one added hole.
This finding can be substantiated from an estimate of the electronic energy 
of an isolated $2\times2$ plaquette.
One obtains the eigenvalue structure $-2t-t'; t'; t'; 2t-t'$ with the 
last level unoccupied due to the presence of the hole. 
This yields the energy per site
$E-E_{AFM}=J\alpha - t(1-t'/2t)/8$ where $-J\alpha$ is the energy per
bond of the AFM phase. Additional carriers have to overcome a large gap ($2t-2t'$) leading to a rapid 
rise of the energy (cf. Fig.~\ref{edx}). At  $n_h=1/16$ other configurations
take a smaller advantage from $t'<0$ while 
the magnetic energy cost is higher in the checkerboard.
Therefore a substantial value of $|t'/t|$ is needed to
stabilize checkerboards with respect to stripes and polarons.

At higher doping and for large $|t'|$ it pays to break more AFM bonds
 to have larger ferromagnetic islands to accommodate the holes.
In this sense the SFC/LFC states locally reflect the tendency
of the extended Hubbard model towards ferromagnetism for large $|t'|$
\cite{toh94,han97} and can be seen as inhomogeneous precursors of the uniform
 ferromagnetic phase, although they do not have a net ferromagnetic
 moment.  

%%%%%%%%%%%%%%%%%%%%%%%%%%%%%%%%%%%%%%%%%%%%%%%%%%%%%%%%%%%%%%%%%%%%%%
%%%%%%%%%%%%%%%%%%%%%%%%%%%%%%%%%%%%%%%%%%%%%%%%%%%%%%%%%%%%%%%%%%%%%%%%3

At intermediate dopings $n_h\sim0.1$ and $t'/t=-0.5$ a phase separated 
solution between the  SFC and LFC has lower energy than $d=4$ stripes.
Ignoring the long-range Coulomb interaction
the energy of this solution is given by the Maxwell construction  [dashed
black line in Fig.~\ref{edx}(b)]. 
This solution will be frustrated in the presence of the long-range Coulomb 
interaction giving rise to a mesoscopic phase
separation\cite{lor01Ilor02ort06}.  
However, since the solutions have the same symmetry they can be mixed
at a quite small length scale which implies a low Coulomb cost. 

At doping $n_h<1/16$ the SFC is expected to phase separate with the AFM
solution. MC for this case is also shown in Fig.~\ref{edx}(b) 
while the corresponding charge- and spin distribution for $n_h=0.047$ is 
shown in Fig.~\ref{spincanting}(b).
In this case the two phases are quite different which can imply a high
energy cost for mesoscopic phase separation\cite{lor01Ilor02ort06} 
since a large surface energy may be expected at the interface. 
We have found, on
the contrary, that allowing for spin canting one finds a solution  
with negligible surface energy cost as shown by the filled circle at
$n_h=0.047$ in Fig.~\ref{edx}(b). Indeed, although the solutions have 
a substantial amount of interface due to the finiteness of the cluster
[Fig.~\ref{spincanting}(b)], the energy  is very close to the MC
line, for which, the surface energy cost is assumed to be zero. 
This is because the surface energy is dominated by magnetic effects,
but spin twisting allows for a smooth  ``flipping'' 
of the magnetic order parameter from the SFC to the pure AFM in accord
with standard arguments for an order parameter which breaks a 
continuous symmetry\cite{imr75}. 
We also see from  Fig.~\ref{edx}(b) that  for the pure checkerboard solution
at $n_h=1/16$ the spin canting energy gain is negligible.  

Mesoscopic textures like the one shown in Fig.~\ref{spincanting}(b) will
also be influenced by disorder effects which will randomly pin clusters of one
or the other phase according to their charge. It is clear from
Fig.~\ref{spincanting}(b) that this will also induce substantial disorder on
the spin degrees of freedom providing a natural mechanism for spin-glass
effects often seen in underdoped cuprates. 

%%%%%%%%%%%%%%%%%%%%%%%%%%%%%%%%%%%%%%%%%%%%%%%%%%%%%%%%%%%%%%%%%%%%%%
%%%%%%%%%%%%%%%%%%%%%%%%%%%%%%%%%%%%%%%%%%%%%%%%%%%%%%%%%%%%%%%%%%%%%%%%

As in previous studies we find that with a ratio of $|t'/t|$ appropriate 
for  LCO the more stable inhomogeneities  are stripes. As $|t'/t|$
is increased checkerboard structures are favored.
Bi-2212 and Na-CCOC are estimated to lie close to the transition\cite{note}
providing an explanation for the puzzling difference in symmetry
observed. The limited accuracy of specific parameter estimates and of the
critical  $|t'/t|$ for checkerboard structures does not allow to exclude
stripes in the bulk of these materials. Quantum fluctuations may also
blur the difference among different configurations close to the transition.

So far checkerboard textures have only been detected by surface sensitive
probes but are not yet established as a bulk phenomenon.
Since the Hubbard $U$ interaction is screened from its atomic  to
the solid-state value by the polarization of the
environment\cite{bri95mei95} a substantial increase of $U$ is
expected to occur at surfaces. Increasing the value to $U/t=10$ we find
[cf. Fig.\ref{spincanting}(a)] that checkerboard structures are more favored 
with respect to stripes and
the critical value of  $|t'/t|$ is decreased by about 15\%. It is thus
possible that checkerboard structures are an example of electronic 
surface reconstruction and occur only at surfaces\cite{mil04}.

In conclusion, we have shown that the next-nearest neighbor
hopping plays a dominant role in determining the symmetry of CO 
textures in cuprates. Whereas for a ratio $t'/t=-0.2$ as appropriate for
LCO we find stripes as the most stable inhomogeneities, a crossover to  
checkerboard solutions occurs for large but still realistic values 
of $|t'/t|$. Checkerboards behave quite differently as a function of doping
with respect to stripes in
that the periodicity of the modulation is locked as observed
experimentally\cite{han04}. The competition between FM clusters and 
antiphase domains rules the relative stability of the various textures.
Spin canting produces a negligible energy gain for ordered structures
but it is fundamental for nanoscopically mixed phases.    

A change of symmetry from stripes in the bulk of LCO and YBCO to
checkerboards at the surface of Bi-2212 and Na-CCOC appears 
quite naturally, thereby explaining the puzzling lack of universality found for 
 the symmetry of CO. On the other hand CO itself  appears to be quite
ubiquitous in cuprates. 
 
We acknowledge interesting discussions with C. Di Castro. 
M.G. and J.L. acknowledge financial support from the MIUR-PRIN2005
n. 2005022492. G.S. acknowledges hospitality from the IMR Sendai where 
part of this work was carried out.

%%%%%%%%%%%%%%%%%%%%%%%%%%%%%%@@@@@@@@@@@@@@@@@@@@@@@@@@@@@@@@@@@@@@@@@@@@@@@@@@@


\begin{thebibliography}{10}

\bibitem{tra95}
J.~M. Tranquada {\it et al.}, Nature
  (London) {\bf 375},  561  (1995).

\bibitem{yam98} K. Yamada {\it et al.},
                Phys. Rev. B {\bf 57}, 6165 (1998).
\bibitem{abb05}
P. Abbamonte {\it et al.},
  Nature Phys. {\bf 1},  155  (2005).

\bibitem{moo00nat}
H.~A. Mook {\it et al.}, Nature (London) {\bf 404},  729  (2000).

\bibitem{tra98}
J.~M. {Tranquada}, Physica B {\bf 241},  745  (1998).

\bibitem{hin04}
V. Hinkov {\it et al.}, Nature (London) {\bf 430},  650  (2004).

\bibitem{fin04}
B. V. Fine,  Phys. Rev. B {\bf 70},  224508 (2004).

\bibitem{hof02sci}
J.~E. Hoffman {\it et al.}, Science {\bf 295},  466  (2002).

\bibitem{how03}
C. Howald {\it et al.}, Phys.\ Rev.\ B
  {\bf 67},  014533  (2003).

\bibitem{ver04}
M. Vershinin {\it et al.}, Science {\bf
  303},  1995  (2004).

\bibitem{mce05} K. McElroy, {\it et al.}, Phys. Rev. Lett. {\bf 94}, 197005 (2005).

\bibitem{han04}
T. {Hanaguri} {\it et al.}, Nature (London) {\bf 430},  1001  (2004).

\bibitem{she05}
K.~M. Shen {\it et al.},
  Science {\bf 307},  901  (2005). 

\bibitem{sei00} G. Seibold {\it et al.}, Eur. Phys. J. B {\bf 13}, 87 (2000).


\bibitem{cotheory}
J. Zaanen and O. Gunnarsson, Phys. Rev. B {\bf 40},  7391  (1989);
K. Machida, Physica C {\bf 158},  192  (1989);
H.~J. Schulz, Phys. Rev. Lett. {\bf 64},  1445  (1990);
D. Poilblanc and T.~M. Rice, Phys. Rev. B {\bf 39},  9749  (1989);
U. L\"ow {\it et al.}, Phys. Rev. Lett.
  {\bf 72},  1918  (1994);
C. Castellani, C. Di Castro, and M. Grilli, Phys. Rev. Lett. {\bf 75}, 4650 (1995).

\bibitem{wak04wak99}
S. Wakimoto {\it et al.},
  Phys.\ Rev.\ Lett. {\bf 92},  217004  (2004);
Phys.\ Rev.\ B {\bf 60},  769  (1999).

\bibitem{pav01}
E. Pavarini {\it et al.},
  Phys.\ Rev.\ Lett. {\bf 87},  047003  (2001).

\bibitem{tan04}
K. Tanaka {\it et al.}, Phys.\ Rev.\ B {\bf 70},  092503  (2004).

\bibitem{sei98}
G. Seibold, Phys. \ Rev. \ B {\bf 58},  15520  (1998).

\bibitem{lor02bsei04a}
J. Lorenzana and G. Seibold, Phys.\ Rev.\ Lett. {\bf 89},  136401  (2002);
G. Seibold and J. Lorenzana, Phys.\ Rev.\ B {\bf 69},  134513  (2004).

\bibitem{tra04}
J.~M. Tranquada {\it et al.}, Nature (London) {\bf 429},  534  (2004).

\bibitem{uch91}
S. Uchida {\it et al.}, Phys.\ Rev.\
  B {\bf 43},  7942  (1991).


\bibitem{lor03}
J. Lorenzana and G. Seibold, Phys.\ Rev.\ Lett. {\bf 90},  066404  (2003).

\bibitem{sei05lor05sei06}
G. Seibold and J. Lorenzana, Phys.\ Rev.\ Lett. {\bf 94},  107006
(2005); Phys.\ Rev.\ B {\bf 73},  144515  (2006);
 J. Lorenzana, G. Seibold, and R. Coldea, Phys.\ Rev.\ B {\bf 72},  224511
  (2005).


\bibitem{whi03}
S.~R. White and D.~J. Scalapino, Phys.\ Rev.\ Lett. {\bf 91},  136403  (2003).

\bibitem{toh94}
T. Tohyama and S. Maekawa, Phys.\ Rev.\ B {\bf 49},  3596  (1994).

\bibitem{han97}
T. Hanisch, E. M\"uller-Hartmann, and G.~S. Uhrig, Phys. Rev. B {\bf 56},
  13960  (1997).

\bibitem{lor01Ilor02ort06}
J. Lorenzana, C. Castellani, and C. {Di Castro}, Phys.\ Rev.\ B {\bf 64},
  235127  (2001); Europhys. Lett. {\bf 57},  704
   (2002); C. Ortix , J. Lorenzana and C. {Di Castro}, Phys.\ Rev.\ B
   in press. 


\bibitem{imr75}
Y. Imry and S. Keng Ma, Phys. Rev. Lett. {\bf 35},  1399  (1975).

\bibitem{note} The case of YBCO is more difficult because $|t'/t|$ is
  probably larger than in LCO, but orthorombicity is more pronounced and requires
a more specific calculation.

\bibitem{bri95mei95}
J. {van den Brink} {\it et al.},
  Phys.\ Rev.\ Lett. {\bf 75},  4658  (1995);
M.~B.~J. Meinders {\it et al.}, Phys.\  Rev.\ B {\bf 52},  2484  (1995).


\bibitem{mil04}
S. {Okamoto} and A.~J. {Millis}, Nature (London) {\bf 428},  630  (2004).

\end{thebibliography}
\end{document}